\documentstyle[epsfig,12pt]{article}
\def \bea{\begin{eqnarray}}
\def \beq{\begin{equation}}
\def \bo{B^0}
\def \bra#1{\langle #1 |}

\def \eea{\end{eqnarray}}
\def \eeq{\end{equation}}

\def \ket#1{| #1 \rangle}

\def \lp{\lambda_{\pi \pi}}
\def \mat#1#2{\langle #1 | #2 \rangle}
\def \ob{\overline{B}^0}

\def \pr{\parallel}

\begin{document}
\Large
\centerline {\bf Information on $B \to \pi \pi$ Provided by}
\centerline{\bf the Semileptonic Process $B \to \pi \ell \nu$
\footnote{Enrico Fermi Institute preprint EFI 01-28, hep-ph/0108024.
Submitted to Physical Review D.}}
\normalsize
 
\vskip 2.0cm
\centerline {Zumin Luo~\footnote{zuminluo@midway.uchicago.edu} and
Jonathan L. Rosner~\footnote{rosner@hep.uchicago.edu}}
\centerline {\it Enrico Fermi Institute and Department of Physics}
\centerline{\it University of Chicago, 5640 S. Ellis Avenue, Chicago, IL 60637}
\vskip 4.0cm
 
\begin{quote}

Analysis of the present data on the semileptonic process $B \to \pi \ell \nu$
indicates that they have not yet reached the precision to provide adequate
information on the $B \to \pi$ form factor $F_+(q^2)$, which for $q^2 =
m_\pi^2$ is known to be related to the factorized color-favored (``$T$'', or
``tree'') contribution to $B^0 \to \pi^+ \pi^-$.  It is shown here that
with around 500 $B \to \pi \ell \nu$ events in
which rate and spectrum are measured one can improve the accuracy of $T$ by a
significant amount.  A recent CLEO determination of
the $D^* D \pi$ coupling constant is compared with an earlier prediction,
and its role in the description of the $B \to \pi$ form factors is noted.
When combined with an estimate of the penguin amplitude (``$P$'') obtained
using flavor SU(3) symmetry from $B \to K \pi$ decays, information  on $T$
allows one to gauge the effects of the penguin amplitude on
extraction of the weak phase $\alpha = \phi_2$ from the time-dependent
CP-violating rate asymmetry in $B^0 \to \pi^+ \pi^-$.  The constraint on 
$\alpha$ implied by a recent experimental result on this asymmetry is described.

\end{quote}
\bigskip

\noindent
PACS Categories: 13.25.Hw, 14.40.Nd, 14.65.Fy, 11.30.Er

\vfill
\newpage

\section{Introduction}

The semileptonic process $B \to \pi \ell \nu$ is known to provide information 
on the $B \to \pi$ form factor $F_+(q^2)$, which for $q^2 = m_\pi^2$ is related to the factorized color-favored (``$T$'', or ``tree'') contribution to $B^0 \to
\pi^+ \pi^-$. In the present paper we show that while present semileptonic data
have not yet reached adequate precision, with around 500 $B \to \pi \ell \nu$
events in which rate and spectrum are measured one can improve the accuracy of
$T$ by a significant amount.  We then discuss the benefits of such a
determination.
  
A connection between the decays $B^0 \to \pi^- \ell^+ \nu_l$ and $B^0 \to
\pi^+ \pi^-$ was noted some time ago by Voloshin \cite{Vol}, who 
derived the relation
\beq \label{eqn:Vol}
\frac{\Gamma(B^0 \to \pi^- e^+ \nu_e)}{\Gamma(B^0 \to \pi^+ \pi^-)}
= \frac{M_B^2}{12 \pi^2 f_\pi^2} \simeq 13.7~~~(f_\pi = 131~{\rm MeV})~~~,
\eeq
using a pole model for the $B \to \pi$ form factor $F_+(q^2)$.  This relation
assumes the dominance of a ``tree'' ($T$) contribution to $B^0 \to \pi^+ \pi^-$
in the notation of Ref.\ \cite{GHLR}.  The CLEO \cite{CLEOSL} and Belle
\cite{BelleSL} Collaborations have measured the branching
ratio for the semileptonic process. Averaging their results yields
\beq \label{eqn:BSL}
{\cal B}(B^0 \to \pi^- e^+ \nu_e) = (1.4 \pm 0.3) \times 10^{-4}~~,
\eeq
while an average of CLEO \cite{CLPP}, Belle \cite{BePP}, and BaBar \cite{BaPP}
($B^0$ and $\ob$-averaged) branching ratios \cite{JRTASI} gives
\beq \label{eqn:ppav}
{\cal B}(B^0 \to \pi^+ \pi^-) = (4.4 \pm 0.9) \times 10^{-6}~~~.
\eeq
The experimental ratio of these two branching ratios is $\Gamma(B^0 \to \pi^-
e^+ \nu_e)/\Gamma(B^0 \to \pi^+ \pi^-) = 32 \pm 9$, a factor of 2.3 above
Eq.~(\ref{eqn:Vol}), which indicates either that the
``tree'' contribution is substantially overestimated in (\ref{eqn:Vol}),
or that some other process is interfering destructively with the tree amplitude
to reduce the $B^0 \to \pi^+ \pi^-$ decay rate.  A prime candidate for this
amplitude is the ``penguin'', or $P$ amplitude in the notation of \cite{GHLR}.
If this amplitude were sufficiently important to reduce the expected $B^0 \to
\pi^+ \pi^-$ rate by roughly a factor of 2.3, it could have important effects
on the extraction of the weak phase $\alpha = \phi_2$ entering the
Cabibbo-Kobayashi-Maskawa (CKM) matrix \cite{penpol}.  This question has now
acquired particular urgency as a result of the first report of results on
CP-violating parameters in $B^0 \to \pi^+ \pi^-$ \cite{Bapipi}.

Many attempts have been made to use data to estimate the ``penguin pollution''
of the $B^0 \to \pi^+ \pi^-$ amplitude, including an isospin analysis requiring
the measurement of $B^0 \to \pi^0 \pi^0$ and $B^+ \to \pi^+ \pi^0$ decays
\cite{GL} (we assume charge-conjugate processes are measured when required),
methods which use only a partial subset of the above information
\cite{GQChGLSS}, and numerous methods based on flavor SU(3)
\cite{GHLR,SilWoSU}.  Earlier data hinted that the penguin amplitude was
interfering destructively with the tree in $B^0 \to \pi^+ \pi^-$
\cite{GRdest,Hdest}.

In the present paper we describe measurements of $B^0 \to \pi^- e^+ \nu_e$
decays which can significantly improve information on the magnitude of the
tree ($T$) contribution to $B^0 \to \pi^+ \pi^-$.  Such an improvement is
needed to tell whether tree and penguin amplitudes are really interfering
destructively in $B^0 \to \pi^+ \pi^-$.  We discuss the role of the $B^*$ pole
in this process, whose contribution is related through heavy quark symmetry to
a recent CLEO measurement of the $D^* D \pi$ coupling constant \cite{CLg}.
We then show how information on $T$ helps to determine the weak phase $\alpha$
using limits on CP violation in $B^0 \to \pi^+ \pi^-$.

Our approach differs from that advocated in Refs.\ \cite{GRdest,GRVP,Neubert},
in which the tree amplitude is estimated from the rate
for $B^+ \to \pi^+ \pi^0$.  In that process, there is an additional
color-suppressed amplitude (called $C$ in the language of Ref.\ \cite{GHLR}),
whose magnitude and phase with respect to $T$ cannot be independently estimated
using present data but must be calculated.  One then has $A(B^+ \to \pi^+
\pi^0) = -(T+C)/\sqrt{2}$, and with $C \simeq 0.1T$, one arrives at estimates
rather similar to those in the present paper.  (The $C$ amplitude was neglected
altogether in Ref.~\cite{GRdest}.) The semileptonic process avoids
dependence on the theoretical calculation of $C/T$.

In Section II we give some basic expressions for the $B^0 \to \pi^- e^+
\nu_e$ and $B^0 \to \pi^+ \pi^-$ decays.  Information
on the $B \to \pi$ form factors is reviewed in Section III.  The
$D^* D \pi$ measurement and its implications for the $B^* B \pi$ coupling
and the $B^*$ pole in the $B \to \pi$ form factor are described in
Section IV.  We then bracket the possible magnitudes of the tree amplitude $T$
depending on measurements of the spectrum in $B^0 \to \pi^- e^+ \nu_e$
(Section V).  The extraction of the penguin amplitude from $B \to K \pi$
decays with the help of flavor SU(3) allows us to determine the
extent to which $P$ and $T$ are interfering destructively in $B^0 \to
\pi^+ \pi^-$, and hence to determine the correction to the weak phase
$\alpha$ which is needed when extracting it from CP-violating
asymmetries in that process (Section VI).  We summarize in Section VII.

\section{Semileptonic and nonleptonic tree decays}

For a generic heavy-to-light decay $H \to \pi$, the non-perturbative
matrix element is parametrized by two independent form factors:
\beq
\bra{\pi(p)} \bar{u}\gamma_{\mu}b \ket{H(p+q)} = 
\left(2p+q-q\frac{m_H^2-m_{\pi}^2}{q^2}\right)_{\mu}F_+(q^2) + 
q_{\mu}\frac{m_H^2-m_{\pi}^2}{q^2}F_0(q^2)~ ,
\eeq
with $H$ being a $B$ or $D$ pseudoscalar meson. The subscript $H$ has
been suppressed in the two form factors. In the case of massless
leptons
(which is an excellent approximation for $\ell=e, \mu$), only $F_+(q^2)$
contributes to the differential decay rate
\beq \label{eqn:diff}
\frac{d\Gamma}{dq^2}(H^0 \to \pi^-\ell^+ \nu_{\ell}) = 
\frac{G_F^2|V_{qQ}|^2}{24\pi^3}|\vec{p}_{\pi}|^3|F_+(q^2)|^2~~ ,
\eeq
where $V_{qQ}$ is the relevant CKM matrix element. We will take
$|V_{cd}|=0.224 \pm 0.016$ and $|V_{ub}|=0.0036 \pm 0.0010$ from
Ref.~\cite{PDG}. To obtain the total width, one should integrate Eq.\
(\ref{eqn:diff}) over the entire physical region, \mbox{$0 \le q^2 \le
(m_H-m_{\pi})^2$}, which requires the precise knowledge of the
normalization [i.e., $F_+(0)$] and $q^2$ dependence of the form factor.

\begin{figure}[t]
\centerline{\epsfysize = 2in \epsffile{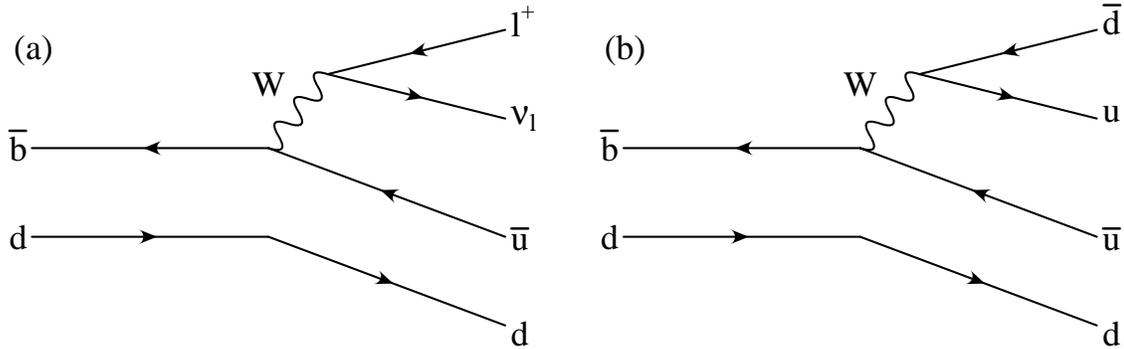}}
\caption{Feynman diagrams for semileptonic and nonleptonic tree decays of 
a $B^0$ meson.
\label{fig:utrees}}
\end{figure}

The lepton pair can be replaced with a pion, as shown in
Fig.~\ref{fig:utrees} for the decay of a $B^0$ meson. The resulted diagram
is the ``tree'' contribution to
the nonleptonic decay $B^0 \to \pi^+ \pi^-$. In the limit of small
$m_{\pi}$, the two diagrams in Fig.~\ref{fig:utrees} are related by the
Bjorken relation \cite{BJ}
\beq \label{eqn:pipi}
\Gamma_{\mathrm{tree}}(B^0 \to \pi^+
\pi^-)=6\pi^2f_{\pi}^2|V_{ud}|^2|a_1|^2\left.\frac{d\Gamma(B^0 \to \pi^-
\ell^+ \nu_{\ell})}{dq^2}\right|_{q^2=m_{\pi}^2} .
\eeq
where $|a_1|$ is the QCD correction. We shall take $|a_1|=1.0$, which is
a sufficiently good approximation for our present purpose.

\section{$H \to \pi$ form factors}

In the absence of a spectrum measurement, one cannot directly employ
Eq.~(\ref{eqn:pipi}) to calculate $T$.  Present extraction of $T$ using this
relation relies on assumptions of particular form factor shapes.
One can test such assumptions using data on the $B^* B \pi$ coupling
extracted using heavy quark symmetry from the corresponding $D^* D \pi$
coupling, and using present information from lattice gauge theories.
Form factors parametrized in a manner consistent with such constraints can then
be used to anticipate the number of events necessary to extract $T$ from
(\ref{eqn:pipi}) in a model-independent way.

Lacking experimental measurements of the form factors $F_+(q^2)$ and
$F_0(q^2)$, people have proposed \cite{models} several models to describe
their behavior, among which is the single-pole model:
\beq \label{eqn:monopole}
F_+(q^2)=\frac{f_{H^*}}{2m_{H^*}}\frac{g_{H^*H\pi}}{1-q^2/m_{H^*}^2}~~ ,
\eeq
where we adopt the following convention:
\bea
\bra{0} V_{\mu} \ket{H^*(p,\epsilon)} & = & f_{H^*}m_{H^*}\epsilon_{\mu}~,\\
\mat{H^-(p)\pi^+(q)}{H^{*0}(p+q,\epsilon)} & = & g_{H^*H\pi}(q\cdot\epsilon)~.
\eea  
However, this form factor gives total widths of \mbox{$D^0 \to \pi^-
\ell^+ \nu_{\ell}$} and \mbox{$B^0 \to \pi^- \ell^+
\nu_{\ell}$} which are both larger than the experimental
values, as will be shown in Section IV. So the monopole form factors are
not enough to describe the physics involved in the $H \to \pi$ decays.
 
Multipole form factors naturally become our next choice.
On the basis of lattice gauge theory calculations,
Becirevic and Kaidalov \cite{BK} proposed a simple parametrization
which is essentially a dipole for $F_+(q^2)$,
\bea
F_+(q^2) &=& \frac{c_H(1-\alpha_H)}{(1-q^2/m_{H^*}^2)(1-\alpha_H
q^2/m_{H^*}^2)}~~, \label{eqn:dipole}\\
F_0(q^2) &=& \frac{c_H(1-\alpha_H)}{1-q^2/(\beta_H m_{H^*}^2)}~~.
\eea
In the infinite quark mass limit, the quantities $(c_H\sqrt{m_H},
(1-\alpha_H)m_H, (\beta_H-1)m_H)$ should scale as constants. $c_H$ is
related to the coupling constant $g_{H^*H\pi}$ as
\beq
c_H = \frac{f_{H^*}g_{H^*H\pi}}{2m_{H^*}}~~ .
\eeq
This parametrization has enough freedom to describe lattice results, which
typically are obtained for values of $q^2$ above about 13 GeV$^2$
\cite{BK,Abada,FermLat}.  We shall employ it to judge the statistical accuracy
needed in extrapolating the $B \to \pi \ell \nu$ spectrum to
$q^2 = m_\pi^2$, where the Bjorken factorization relation (\ref{eqn:pipi})
provides an estimate of $T$.  A similar problem arises when one wishes to
extrapolate to the zero-recoil limit in estimating the CKM matrix element
$|V_{cb}|$ from the exclusive process $B \to D^{(*)} \ell \nu$, since both
the normalization and shape of the spectrum have to be determined.

It should be pointed out that $f_{D^*}$, $f_{B^*}$ and $g_{B^*B\pi}$ are
far from being determined, though $g_{D^*D\pi}$ has been measured \cite{CLg}. 
Very different values of $f_{D^*}$ and $f_{B^*}$ have been obtained on the
lattice and in various models (see Table \ref{tab:decayconstants}
\cite{Becirevic,Bowler,Hwang,Wang,Huang}).  We will discuss $g_{B^*B\pi}$ in
Section IV.

\begin{table}
\caption{Vector meson decay constants (MeV) from different
calculations. \label{tab:decayconstants}}
\begin{center}
\begin{tabular}{l c c} \hline \hline
	& $f_{D^*}$ & $f_{B^*}$ \\ \hline
Becirevic {\it et al.} \cite{Becirevic} & $245 \pm 20 ^{+3}_{-2}$ & $196
\pm 24 ^{+39}_{-2}$ \\
UKQCD \cite{Bowler} & $268 ^{+32}_{-40}$ & $236 ^{+45}_{-39}$ \\
Hwang \& Kim \cite{Hwang} & $327 \pm 13$ & $252 \pm 10$ \\
Wang \& Wu \cite{Wang} & $354 \pm 90$ & $206 \pm 39$ \\
Huang \& Luo \cite{Huang} & & $190 \pm 30$ \\ \hline \hline
\end{tabular}
\end{center}
\end{table}

\section{Implications of $g_{D^* D \pi}$ measurement}

We now describe the CLEO measurement of the $D^* D \pi$ coupling constant
\cite{CLg} and review its significance in the light of earlier predictions
\cite{JRD,Amun,Drev}.  The observed value of the total $D^{*+}$ width is
$\Gamma(D^{*+}) = (96 \pm 4 \pm 22)$ keV, in satisfactory agreement with a
prediction of 84 keV made some time ago by comparison with $K^* \to
K \pi$ and $K^* \to K \gamma$ decays \cite{JRD}.  Other predictions of
\cite{JRD} are compared with the current experimental situation
\cite{PDG} in Table \ref{tab:Dstars}.  The agreement is not bad, and can
be improved by assuming about a 30\% increase in the absolute square of
the matrix element for the magnetic dipole transitions $D^* \to D \gamma$
with respect to the value in Refs.\ \cite{JRD}.  The experimental
branching ratios at the time of these predictions differed from them much
more significantly.

\begin{table}
\caption{Predictions for decays $D^* \to D \pi$ and $D^* \to D \gamma$ based
on comparison with $K^* \to K \pi$ and $K^* \to K \gamma$ decays.
\label{tab:Dstars}}
\begin{center}
\begin{tabular}{r r r c} \hline \hline
      &  \multicolumn{2}{c}{Predicted}  & Experimental \\
      & Partial Width & Branching Ratio & Branching Ratio \\
Decay &     (keV)     &      (\%)       &      (\%)       \\ \hline
$D^{*+} \to D^+ \pi^0$ & 25.9 & 30.9 & $30.7 \pm 0.5$ \\
   $ \to D^0 \pi^+$    & 56.9 & 67.8 & $67.7 \pm 0.5$ \\
   $ \to D^+ \gamma$   & \underline{~~1.1} & 1.3 & $1.6 \pm 0.4$ \\
                       & 83.9 & & \\ \hline
$D^{*0} \to D^0 \pi^0$ & 39.7 & 70.6 & $61.9 \pm 2.9$ \\
   $ \to D^0 \gamma$   & \underline{16.5} & 29.4 & $38.1 \pm 2.9$ \\
                       & 56.2 & & \\ \hline \hline
\end{tabular}
\end{center}
\end{table}

\begin{figure}
\centerline{\epsfysize = 4.3in \epsffile{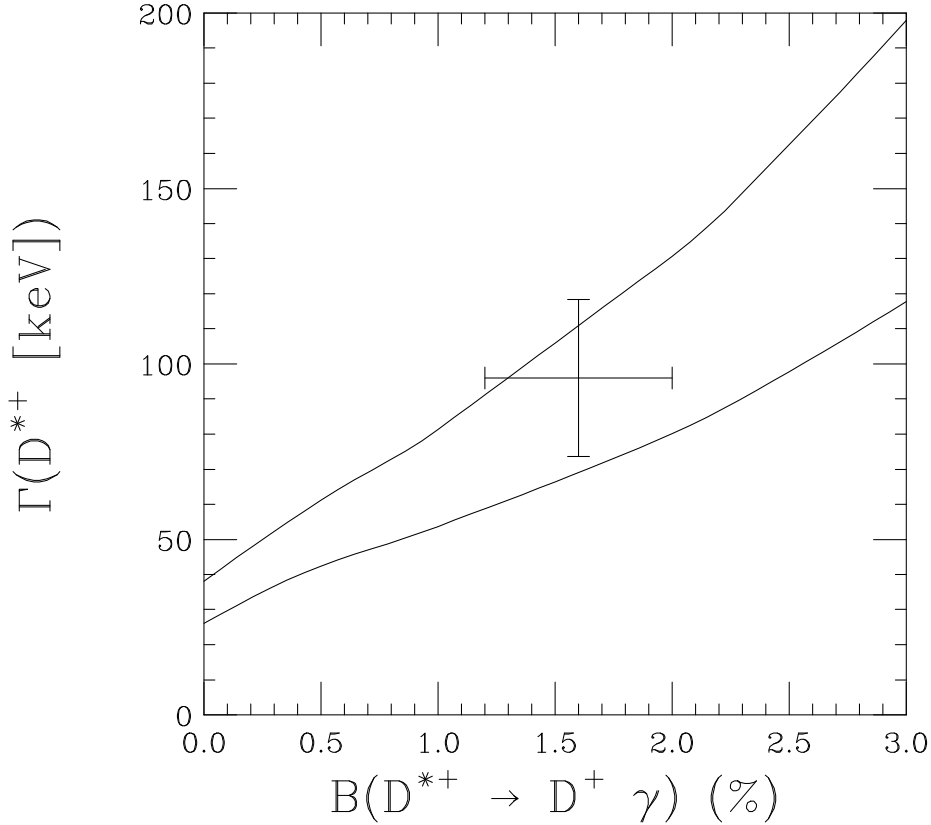}}
\caption{Prediction of Ref.\ \cite{Amun} for $\Gamma(D^{*+})$ as a function of
the branching ratio for $D^{*+} \to D^+ \gamma$, including leading
SU(3)-breaking effects.  Lines show predicted bounds. The plotted point shows
current data \cite{CLg,PDG}. \label{fig:correl}}
\end{figure}

A more detailed set of calculations was performed on the basis of
chiral and heavy quark symmetry \cite{Amun}, taking into account SU(3)
violating contributions of order $m_q^{1/2}$.  The experimental values are
consistent with the predicted correlation between ${\cal B}(D^{*+} \to D^+
\gamma)$ and $\Gamma(D^{*+})$, as shown in Fig.\ \ref{fig:correl}.

The observed $D^{*+}$ width can be related to a dimensionless $D^* D \pi$
coupling constant $\hat g$ by the expression \cite{Amun,Ruckl}
\beq
\Gamma(D^{*+} \to D^0 \pi^+) = \frac{\hat g^2}{6 \pi f_\pi^2}
|\vec{p}_\pi|^3~~,
\eeq
where $f_\pi = 131$ MeV and $|\vec{p}_\pi| = 39$ MeV/$c$.  Using the
branching
ratio in Table \ref{tab:Dstars} we find $\Gamma(D^{*+} \to D^0 \pi^+) =
65 \pm 15$ keV and $\hat g = 0.59 \pm 0.07$. Therefore
\beq
g_{D^*D\pi}=\frac{2m_D^s}{f_{\pi}}\hat{g}=17.8 \pm 2.1~~,
\eeq
where $m_D^s=1973$~MeV is the spin-averaged mass of the $D^{(*)}$ 
meson. Taking this value of $g_{D^*D\pi}$ and $f_{D^*}=200$~MeV (which
is more than $1\sigma$ smaller than any determination in Table 
\ref{tab:decayconstants}), we get ${\cal B}(D^0 \to \pi^- e^+ \nu_e) =
(4.9 \pm 1.2) \times 10^{-3}$, still larger than the experimental value
$(3.7 \pm 0.6) \times 10^{-3}$ \cite{PDG}. Higher values of $f_{D^*}$
yield even larger branching ratios.

Heavy quark symmetry (HQS) predicts
\beq
g_{B^*B\pi} = \frac{2m_B^s}{f_{\pi}}\hat{g} = 47.9 \pm 5.7~~.
\eeq
Again, even if we take a comparatively small value of $f_{B^*}$ (=160
MeV) and assume a large (e.g., 40\%) violation of HQS (so that
$g_{B^*B\pi}$ can be as small as 29.0), we will get a branching ratio
${\cal B}(B^0 \to \pi^- e^+ \nu_e) = (2.6 \pm 1.4) \times 10^{-4}$ which
is still larger than Eq.~(\ref{eqn:BSL}). Thus we are justified to suspect
the single pole form factor (\ref{eqn:monopole}).

\section{Information on $T$ from semileptonic decays}

The Bjorken relation (\ref{eqn:pipi}) establishes a useful connection between
the semileptonic decays and the nonleptonic ``tree'' decays. Ideally,
$d\Gamma(B^0 \to \pi^- \ell^+ \nu_{\ell})/dq^2$ at $q^2 = m_{\pi}^2$
provides the ``tree'' contribution to the branching ratio for $B^0
\to \pi^+ \pi^-$ (aside from QCD corrections, which have been
found to be a few percent in related processes).  However, in practice one
must measure the semileptonic decay spectrum over a range of $q^2$ in order
to accumulate a sufficient number of events, and therefore must model the
spectrum shape, as in extracting $|V_{cb}|$ from the spectrum for
$B \to D^{(*)} \ell \nu$.

\begin{figure}
\centerline{\epsfysize = 3.5in \epsffile{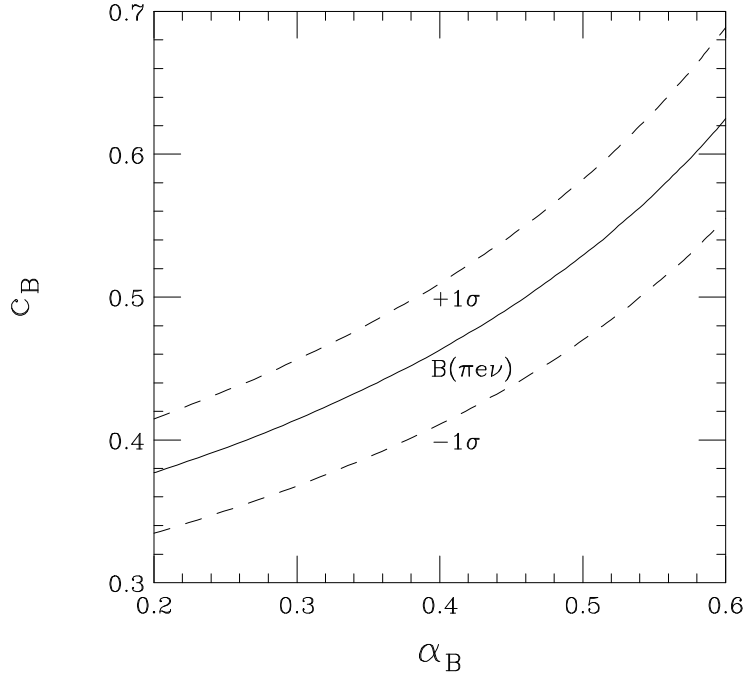}}
\caption{The dependence of $c_B$ on $\alpha_B$ for given values of ${\cal
B}(B^0 \to \pi^- e^+ \nu_e)$. Solid line: ${\cal B}(B^0 \to \pi^- e^+
\nu_e)=1.4 \times 10^{-4}$; upper dashed line: ${\cal B}(B^0 \to \pi^-
e^+ \nu_e)=1.7 \times 10^{-4}$; lower dashed line: ${\cal B}(B^0 \to
\pi^- e^+ \nu_e)=1.1 \times 10^{-4}$.
\label{fig:alpha_c}}
\end{figure}

\begin{figure}
\centerline{\epsfysize = 3.5in \epsffile{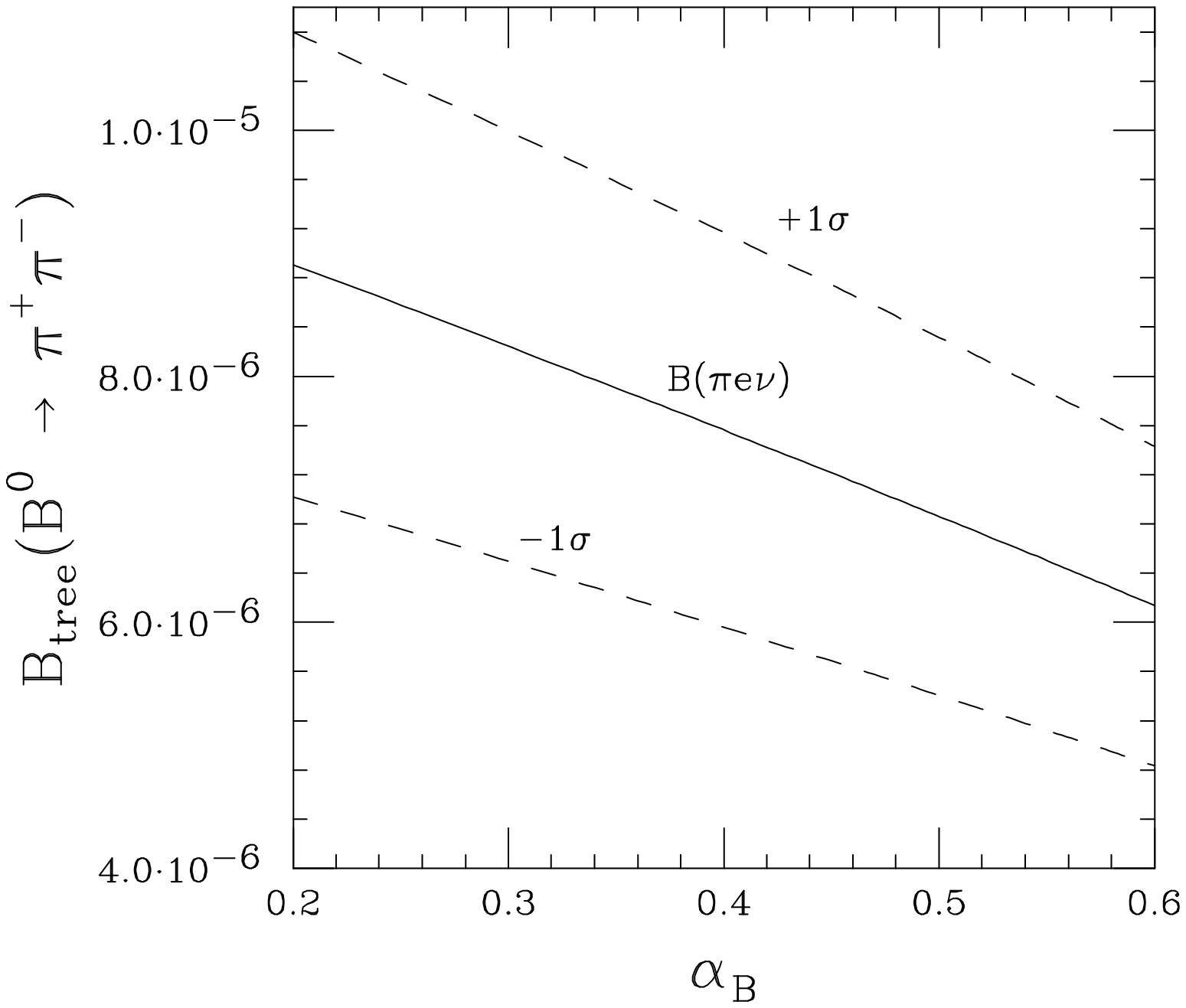}}
\caption{The dependence of ${\cal B}_{\mathrm{tree}}(B^0 \to \pi^+ \pi^-)$
on $\alpha_B$ for given values of ${\cal B}(B^0 \to \pi^- e^+
\nu_e)$. Solid line: ${\cal B}(B^0 \to \pi^- e^+ \nu_e)=1.4 \times
10^{-4}$; upper dashed line: ${\cal B}(B^0 \to \pi^- e^+ \nu_e)=1.7
\times 10^{-4}$; lower dashed line: ${\cal B}(B^0 \to \pi^- e^+
\nu_e)=1.1 \times 10^{-4}$.
\label{fig:alpha_br}}
\end{figure}

The dipole form factor has enough parameters to allow
modeling both a normalization and a spectrum shape.  We use it to gain an idea
of the statistical requirements for a useful spectrum measurement.
The experimental branching ratio (\ref{eqn:BSL}) for the semileptonic
decay $B^0 \to \pi^- e^+ \nu_e$ puts a strong constraint on the dipole
parameters $c_B$ and $\alpha_B$, as shown in Fig.~\ref{fig:alpha_c}.
Accordingly, the ``tree'' branching ratio for $B^0 \to \pi^+ \pi^-$ is
constrained to lie in a certain range (Fig.~\ref{fig:alpha_br}).
It should be noted that Fig.~\ref{fig:alpha_br} does not depend on $|V_{ub}|$,
though Fig.~\ref{fig:alpha_c} can be
altered by any change in $|V_{ub}|$. We can always combine $|V_{ub}|$ with
$c_B$ and view $|V_{ub}|c_B$ as a single parameter. This observation
plays an important role in estimating $T$ from Fig.~\ref{fig:alpha_br}.

\begin{figure}
\centerline{\epsfysize = 4in \epsffile{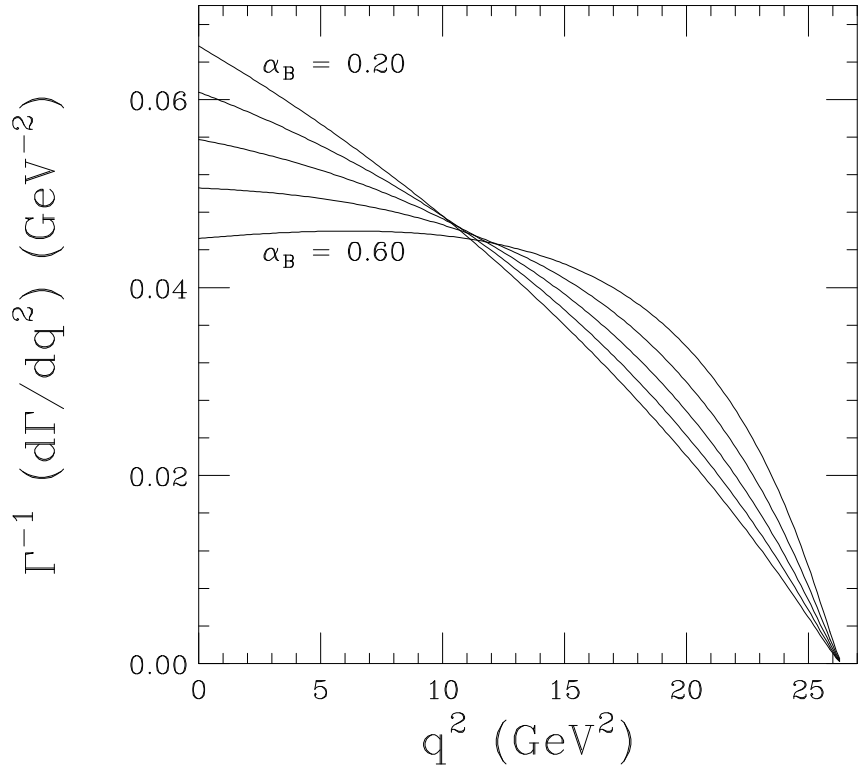}}
\caption{Normalized spectrum of $B^0 \to \pi^- \ell^+
\nu_{\ell}$ for various values of $\alpha_B$. At low $q^2$, the
curves correspond to $\alpha_B=0.20, 0.30, 0.40, 0.50, 0.60$,
from top to bottom. \label{fig:spectrum1}}
\end{figure}

To determine $\alpha_B$ and hence $c_B$ and ${\cal B}_{\mathrm{tree}}(B^0
\to \pi^+ \pi^-)$, one can measure the normalized spectrum
($\frac{1}{\Gamma}\frac{d\Gamma}{dq^2}$) for $B^0 \to \pi^- \ell^+
\nu_{\ell}$. Note that $\frac{1}{\Gamma}\frac{d\Gamma}{dq^2}$ is
independent of $c_B$ and $|V_{ub}|$. Thus measuring its dependence on
$q^2$ will give us very clean information about $\alpha_B$.
Fig.~\ref{fig:spectrum1} shows us that a comparison of the spectrum in
the interval $0 \le q^2 \le 11$ GeV$^2$ with that for $11 \le q^2 \le
26$ GeV$^2$ should be useful in determining $\alpha_B$.

A recent lattice calculation \cite{Abada} obtains values of $\alpha_B$
ranging from about 0.2 to 0.6, $c_B$ from about 0.3 to 0.6, and $F_+(0)$
around $0.27$ with a 25\% error.  
A more recent analysis \cite{Ball} from QCD sum rules on the light-cone
obtains $F_+(0)=0.26 \pm 0.08$, in good agreement with the lattice
result.  This implies that parameters are
within the ranges quoted in Figs.~\ref{fig:alpha_c}-\ref{fig:spectrum1},
and leads to values of ${\cal B}_{\rm tree}(B^0 \to \pi^+ \pi^-)$ ranging
between about $4.5 \times 10^{-6}$ and $11 \times 10^{-6}$, as in
Fig.~\ref{fig:alpha_br}. 

Given the central value of ${\cal B}(B \to \pi \ell \nu)$,
Fig.~\ref{fig:alpha_br} implies that an error $\Delta \alpha_B = 0.1$ will
correspond to an error in $\Delta {\cal B}_{\mathrm{tree}}(B^0 \to \pi^+
\pi^-)$ of about 10\%, or an error in $T$ of about 5\%.  An additional
error will be associated with the statistical error associated with ${\cal
B}(B \to \pi \ell \nu)$ itself.  We shall determine the number of events
required to achieve an error of $\Delta \alpha_B = 0.1$, and estimate the
corresponding total error in $T$.

In Table \ref{tab:fracts} we show the fraction $f$ of $B^0 \to \pi^- \ell^+
\nu_{\ell}$ events below $q^2 = 11$ GeV$^2$ as a function of $\alpha_B$.
In order to obtain an error of $\Delta \alpha_B = 0.1$, one has to determine
$f$ to a precision of $\Delta f = 0.023$.  For a
total of $N$ events in the spectrum, the error in $f$ is $\Delta f =
\sqrt{f(1-f)/N}$, which is about $0.5/\sqrt{N}$ for $f$ near 0.5.  Thus, one
needs about $(0.5/0.023)^2 \simeq 470$ $B \to \pi \ell \nu$ events to
achieve this accuracy.  Such a sample will be associated with an error in
the overall $B \to \pi \ell \nu$ rate of $1/\sqrt{470} \simeq
4.6\%$.  When added in quadrature with the 10\% error in ${\cal
B}_{\mathrm{tree}}(B^0 \to \pi^+ \pi^-)$ associated with the spectrum
shape, this leads to an overall error of 11\% in $|T|^2$ or 5.5\% in
$T$.  One will need considerably more than $470/{\cal B}(B \to \pi \ell
\nu) \simeq  3.4\times 10^6$ $B$ decays to obtain a sample of this size,
since the efficiency of reconstructing the semileptonic decay is
small (e.g., slightly under 2\% at Belle \cite{BelleSL}).  The Belle
Collaboration has reported a signal of 107 events on the basis
of 21.2 fb$^{-1}$, but the background (148 events) is larger than the
signal, and the branching ratio is dominated by systematic
error.  Thus a sample of about 4.4 times the present size would be the
minimum needed to achieve the stated goal, with a larger sample required if
background levels are to be reduced.

\begin{table}
\caption{Dependence of the fraction $f$ of $B^0 \to \pi^- \ell^+ \nu_{\ell}$
events below $q^2 = 11$ GeV$^2$ on the parameter $\alpha_B$.
\label{tab:fracts}}
\begin{center}
\begin{tabular}{c c c c c c} \hline \hline
$\alpha_B$ & 0.2 & 0.3 & 0.4 & 0.5 & 0.6 \\ \hline
    $f$            & 0.618  & 0.595 & 0.568 & 0.538 & 0.503 \\ \hline
\hline
\end{tabular}
\end{center}
\end{table}

\section{Information on $P$ and its interference with $T$}

We shall use present and anticipated information on $T$ based on the
methods described in the previous section, and flavor SU(3) \cite{GHLR} to
obtain information on $P$ from the mainly-penguin process $B^+ \to K^0 \pi^+$.
In this manner we shall end up with an estimate $|P/T| = 0.26 \pm 0.08$, to be
compared to the value of $0.259 \pm 0.043 \pm 0.052$ quoted by Beneke
{\it et al.} \cite{BBNS} on the basis of a theoretical calculation.  (The
inclusion of weak annihilation contributions in \cite{BBNS} raises this value
to $0.285 \pm 0.051 \pm 0.057$.)  Improved input data will potentially
reduce the error on this ratio considerably, allowing for an estimate of direct
CP-violating effects in $B^0 \to \pi^+ \pi^-$ with less recourse to theory.
Furthermore, if $|T|$ turns out to be incompatible with
the experimental magnitude of the amplitude $A(B^0 \to \pi^+ \pi^-) = - (T+P)$,
we shall obtain a constraint on the product $\cos \alpha \cos \delta$,
where $\alpha$ is the CKM phase discussed previously and $\delta$ is the
relative strong phase between tree and penguin amplitudes.  Our discussion
will be an updated version of that presented in \cite{GRVP}.

We shall quote all rates in units of ($\bo$ branching ratio $\times
10^6$). Thus, the average (\ref{eqn:ppav}) of $\bo \to \pi^+ \pi^-$
branching ratios implies
\beq \label{eqn:rat}
|T|^2 + |P|^2 - 2 |TP| \cos \alpha \cos \delta = 4.4 \pm 0.9~~~
\eeq
in these units.
With ${\cal B}_{\rm tree}(B^0 \to \pi^+ \pi^-)$ ranging from (4.5 to 11)
$\times 10^{-6}$ we then estimate $|T| = 2.7 \pm 0.6$.  This is identical to
the value obtained \cite{B2Kfact} from $B^+ \to \pi^+ \pi^0$ with additional
assumptions about the color-suppressed amplitude.

The penguin amplitude can be estimated from $B^+ \to K^0 \pi^+$.  The average
of CLEO \cite{CLPP}, Belle \cite{BePP}, and BaBar \cite{BaPP} branching ratios
\cite{JRTASI} gives
\beq
{\cal B}(B^+ \to K^0 \pi^+) = (17.2 \pm 2.4) \times 10^{-6}~~~,
\eeq
leading to $|P'|^2 = (17.2 \pm 2.4)(\tau^0/\tau^+)$, $|P'| = 4.02 \pm 0.28$,
where we use the lifetime ratio $\tau_{B^+}/\tau_{B^0} = 1.068 \pm 0.016$
\cite{Blifes}. Here $P'$ refers to the strangeness-changing $\bar b \to \bar s$
penguin amplitude, which is dominated by the CKM combination $V_{ts} V^*_{tb}$.

We now estimate the strangeness-preserving $\bar b \to \bar d$ amplitude by
assuming it to be dominated by the CKM combination $V_{td} V^*_{tb}$.  This
may induce some uncertainty if the lighter intermediate quarks also play a
role \cite{BuFP}.  (A slightly different definition of $P$ is used by
\cite{BBNS,GR01} and avoids this problem.)  We find
\beq
|P/P'| \simeq \left| \frac{V_{td}}{V_{ts}} \right| = \lambda |1 - \rho - i
\eta| \simeq 0.22(0.80 \pm 0.15)~~,~~~|P| = 0.71 \pm 0.14~~~,
\eeq
where $\lambda$, $\rho$, and $\eta$ are parameters \cite{WP}
describing the hierarchy of CKM matrix elements.  Combining these results,
we find only that $-0.1 \le \cos \alpha \cos \delta \le 1$, so that destructive
interference is possible but not established.  Reduced errors on $|T|$ and
$|P|$ will be needed for a more definitive conclusion.  In particular, given
the present central values, reduction of the error on $|T|^2$ to
11\%, as achievable with 470 $B \to \pi \ell \nu$ events, would allow one
to infer the presence of destructive interference at about the $2.8
\sigma$ level.

With our present estimates of $|P|$ and $|T|$, we then find $|P/T| = 0.26 \pm
0.08$. Errors on this quantity can be decreased by improving the
measurements of the branching ratio for $B \to \pi \ell \nu$, by measuring
its spectrum, and by reducing the error on $|1 - \rho - i \eta|$, which we
have taken to be greater than in some other determinations \cite{B2KCKM}.

The presence of the $P$ amplitude can affect the determination of the
weak phase $\alpha$ using CP-violating asymmetries in $B^0 \to \pi^+ \pi^-$
decays.  The BaBar Collaboration \cite{Bapipi} has recently reported the
first results for this process.  The decay distributions $f_+~(f_-)$
in an asymmetric $e^+ e^-$ collider at the $\Upsilon(4S)$ when the tagging
particle (opposite to the one produced) is a $B^0$ ($\ob$) are given by
\cite{GL}
\beq
f_{\pm}(\Delta t) \simeq e^{-\Delta t/\tau}[ 1 \pm S_{\pi \pi} \sin
(\Delta m_d \Delta t) \mp C_{\pi \pi} \cos (\Delta m_d \Delta t) ]~~~,
\eeq
where
\beq
S_{\pi \pi} \equiv \frac {2 {\rm Im}(\lp)}{1 + |\lp|^2}~~,~~~
C_{\pi \pi} \equiv \frac{1 - |\lp|^2}{1 + |\lp|^2}~~~
\eeq
and
\beq
\lp \equiv e^{-2 i \beta} \frac{A(\ob \to \pi^+ \pi^-)}
{A(\bo \to \pi^+ \pi^-)}~~~.
\eeq
Here
$$
A(\bo \to \pi^+ \pi^-) \simeq -(|T|e^{i \delta_T} e^{i \gamma} +
 |P| e^{i \delta_P} e^{-i \beta})~~~,
$$
\beq
A(\ob \to \pi^+ \pi^-) \simeq -(|T|e^{i \delta_T} e^{- i \gamma} + 
 |P| e^{i \delta_P} e^{i \beta})~~~,
\eeq
where $\delta_T$ and $\delta_P$ are strong phases of the tree and penguin
amplitudes.  To first order in $|P/T|$, using $\beta + \gamma = \pi - \alpha$
and defining $\delta \equiv \delta_P - \delta_T$, we then have
\beq
\lp \simeq e^{2 i \alpha} \left( 1 + 2i \left| \frac{P}{T} \right| e^{i \delta}
\sin \alpha \right)~~~.
\eeq

\begin{figure}
\centerline{\epsfysize = 4.6in \epsffile{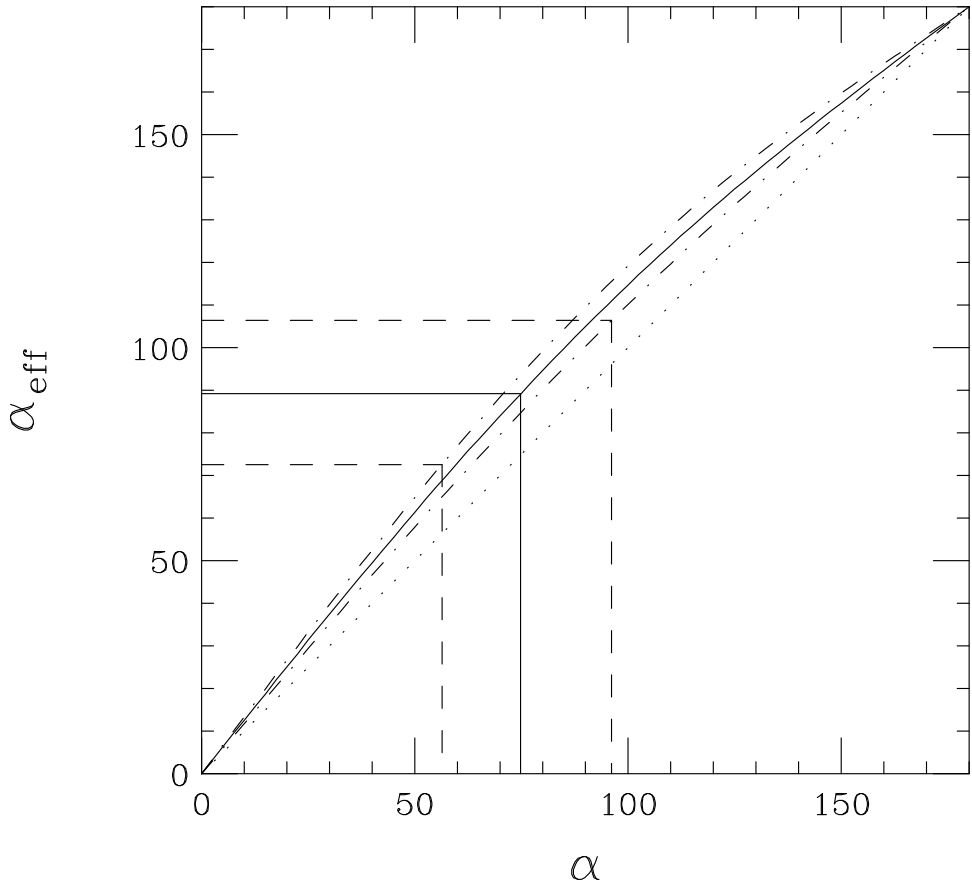}}
\caption{Relation between $\alpha_{\rm eff}$ as measured using $S_{\pi \pi} =
\sin(2 \alpha_{\rm eff})$ and the weak phase $\alpha$ for $|P/T| = 0.26$ and
$\delta = 0$ (solid curve).  Dot-dashed curves correspond to $\pm 1 \sigma$
errors on $|P/T|$.  The dotted line corresponds to $P=0$.  The solid
and dashed lines correspond to the central and $\pm 1 \sigma$ values of $S_{\pi
\pi}$ recently reported by the BaBar Collaboration (allowing also for error
in $|P/T|$).  We show only the range
associated with the region of CKM parameters consistent with other
measurements. \label{fig:alpha}}
\end{figure}

In the limit of small $|P/T|$ and vanishing final-state phase $\delta$, the
$S_{\pi \pi}$ term is just $\sin(2 \alpha_{\rm eff})$, where
\beq
\alpha_{\rm eff} \simeq \alpha + \left| \frac{P}{T} \right| \sin \alpha~~~.
\eeq
A plot of this relation for $|P/T| = 0.26 \pm 0.08$ is shown in Fig.\
\ref{fig:alpha}.  The BaBar Collaboration \cite{Bapipi} has reported $S_{\pi
\pi} = 0.03^{+0.53}_{-0.56} \pm 0.11$ on the basis of 30.4 fb$^{-1}$. The
corresponding central and $\pm 1 \sigma$ values of $\alpha_{\rm eff}$ and
$\alpha$ are shown as the solid and dashed lines on the figure.

To first order in $|P/T|$, the $C_{\pi \pi}$ term may be written
\beq
C_{\pi \pi} \simeq 2 |P/T| \sin \delta \sin \alpha~~~.
\eeq
The BaBar Collaboration's value \cite{Bapipi} $C_{\pi \pi} = -0.25^{+0.45}
_{-0.47} \pm 0.47$ is consistent with zero, as one might expect for a
small final-state phase $\delta$.  This measurement in the future will serve
mainly to constrain $\delta$, given the limited range expected for $|P/T|$
and $\sin \alpha$.  Such a constrained value of $\delta$ will then be useful
in interpreting the flavor-averaged branching ratio (\ref{eqn:ppav}) in terms
of the tree-penguin interference discussed previously.  The combined
measurements of the flavor-averaged $B^0 \to \pi^+ \pi^-$ branching ratio
and the coefficients $S_{\pi \pi}$ and $C_{\pi\pi}$, when combined with
independent determinations of $|T|$ and $|P|$, should allow us to dispense
with the assumptions that the final-state phase $\delta$ is small
and that the weak phase of $P$ is dominated by the top quark in the loop.

\begin{figure}[t]
\centerline{\epsfysize = 4.5in \epsffile{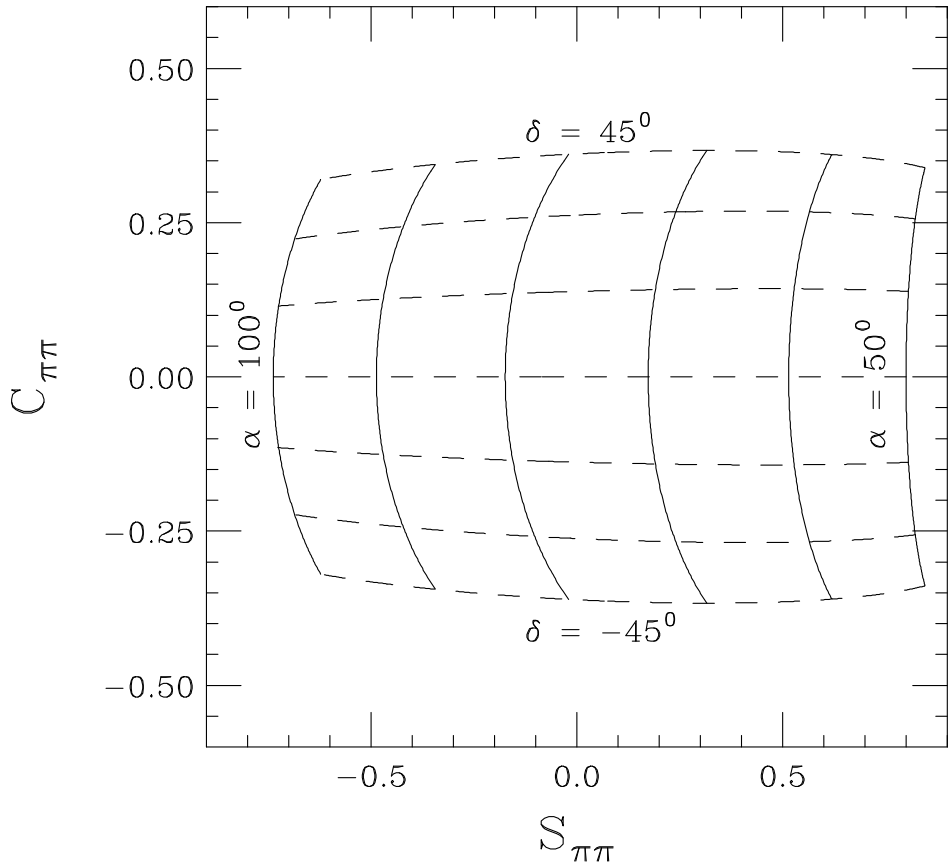}}
\caption{Relation between $S_{\pi \pi}$ and $C_{\pi \pi}$ for fixed values of
$\alpha$ (solid curves) and $\delta$ (dashed curves).  The values of $\alpha$
range in steps of $10^\circ$ from $50^\circ$ (right) to $100^\circ$ (left);
those of $\delta$ range in steps of $15^\circ$ from $-45^\circ$ (bottom)
to $45^\circ$ (top).  Here $|P/T| = 0.26$ has been assumed.
\label{fig:CS}}
\end{figure}

An example is shown in Fig.\ \ref{fig:CS} of how $S_{\pi \pi}$ and $C_{\pi\pi}$
measurements can be used to constrain $\alpha$ and $\delta$.  Values extracted
from such plots can then be checked for consistency with Eq.~(\ref{eqn:rat})
to check our assumption that the phase and magnitude of $P$ is dominated by
the top quark.

\section{Conclusions}

We have discussed rate and spectrum requirements in $B \to \pi \ell \nu_l$
decays needed to reduce errors in the tree-amplitude contribution $T$ to
$B^0 \to \pi^+ \pi^-$.  Better knowledge of $T$ can be combined with an
estimate of the penguin amplitude $P$ to see if destructive tree-penguin
interference is occurring in $B^0 \to \pi^+ \pi^-$, and to evaluate the
correction to the time-dependent CP asymmetry parameters $S_{\pi \pi}$ and
$C_{\pi \pi}$.  Present data lead to the estimate $|P/T| = 0.26 \pm 0.08$ but
substantial improvement will be possible once the semileptonic rate and
spectrum (particularly near $q^2 = 0$) are better measured.  We have estimated
that at least 470 $B \to \pi \ell \nu$ events (about 4.4 times the present
sample size at Belle) are needed to reduce the error on $T$ to 5.5\%. 
For $\alpha$ near $90^\circ$ we predict $\alpha_{\rm eff} - \alpha \simeq (15
\pm 5)^\circ$.  Destructive tree-penguin interference in $B^0 \to \pi^+ \pi^-$
could be significant if $\alpha$ were closer to
the lower limit of about $56^\circ$ allowed by the present analysis.
The form factor $F_+(q^2)$ measured in $B \to \pi \ell \nu_l$ also can be
helpful in estimating
the ``wrong-sign'' amplitude in $B \to D^* \pi$ decays \cite{SCR}.
 
\section*{Acknowledgments}

We thank Martin Beneke, Michael Gronau, Andreas Kronfeld, Zoltan Ligeti,
Harry Lipkin, and Denis Suprun for helpful
suggestions and Aaron Roodman for discussions of experimental capabilities.
Part of this investigation was performed while one of us (J.L.R.) was at the
Aspen Center for Physics.  This work was supported in part by the United
States Department of Energy through Grant No.\ DE FG02 90ER40560.

\def \ajp#1#2#3{Am.\ J. Phys.\ {\bf#1}, #2 (#3)}
\def \apny#1#2#3{Ann.\ Phys.\ (N.Y.) {\bf#1}, #2 (#3)}
\def \app#1#2#3{Acta Phys.\ Polonica {\bf#1}, #2 (#3)}
\def \arnps#1#2#3{Ann.\ Rev.\ Nucl.\ Part.\ Sci.\ {\bf#1}, #2 (#3)}
\def \art{and references therein}
\def \cmts#1#2#3{Comments on Nucl.\ Part.\ Phys.\ {\bf#1}, #2 (#3)}
\def \cn{Collaboration}
\def \cp89{{\it CP Violation,} edited by C. Jarlskog (World Scientific,
Singapore, 1989)}
\def \efi{Enrico Fermi Institute Report No.\ }
\def \epjc#1#2#3{Eur.\ Phys.\ J. C {\bf#1}, #2 (#3)}
\def \f79{{\it Proceedings of the 1979 International Symposium on Lepton and
Photon Interactions at High Energies,} Fermilab, August 23-29, 1979, ed. by
T. B. W. Kirk and H. D. I. Abarbanel (Fermi National Accelerator Laboratory,
Batavia, IL, 1979}
\def \hb87{{\it Proceeding of the 1987 International Symposium on Lepton and
Photon Interactions at High Energies,} Hamburg, 1987, ed. by W. Bartel
and R. R\"uckl (Nucl.\ Phys.\ B, Proc.\ Suppl., vol.\ 3) (North-Holland,
Amsterdam, 1988)}
\def \ib{{\it ibid.}~}
\def \ibj#1#2#3{~{\bf#1}, #2 (#3)}
\def \ichep72{{\it Proceedings of the XVI International Conference on High
Energy Physics}, Chicago and Batavia, Illinois, Sept. 6 -- 13, 1972,
edited by J. D. Jackson, A. Roberts, and R. Donaldson (Fermilab, Batavia,
IL, 1972)}
\def \ijmpa#1#2#3{Int.\ J.\ Mod.\ Phys.\ A {\bf#1}, #2 (#3)}
\def \ite{{\it et al.}}
\def \jhep#1#2#3{JHEP {\bf#1}, #2 (#3)}
\def \jpb#1#2#3{J.\ Phys.\ B {\bf#1}, #2 (#3)}
\def \lg{{\it Proceedings of the XIXth International Symposium on
Lepton and Photon Interactions,} Stanford, California, August 9--14 1999,
edited by J. Jaros and M. Peskin (World Scientific, Singapore, 2000)}
\def \lkl87{{\it Selected Topics in Electroweak Interactions} (Proceedings of
the Second Lake Louise Institute on New Frontiers in Particle Physics, 15 --
21 February, 1987), edited by J. M. Cameron \ite~(World Scientific, Singapore,
1987)}
\def \kdvs#1#2#3{{Kong.\ Danske Vid.\ Selsk., Matt-fys.\ Medd.} {\bf #1},
No.\ #2 (#3)}
\def \ky85{{\it Proceedings of the International Symposium on Lepton and
Photon Interactions at High Energy,} Kyoto, Aug.~19-24, 1985, edited by M.
Konuma and K. Takahashi (Kyoto Univ., Kyoto, 1985)}
\def \lpRoma{XX International Symposium on Lepton and Photon Interactions
at High Energies, Rome, Italy, July 23--27, 2001}
\def \mpla#1#2#3{Mod.\ Phys.\ Lett.\ A {\bf#1}, #2 (#3)}
\def \nat#1#2#3{Nature {\bf#1}, #2 (#3)}
\def \nc#1#2#3{Nuovo Cim.\ {\bf#1}, #2 (#3)}
\def \nima#1#2#3{Nucl.\ Instr.\ Meth. A {\bf#1}, #2 (#3)}
\def \np#1#2#3{Nucl.\ Phys.\ {\bf#1}, #2 (#3)}
\def \npbps#1#2#3{Nucl.\ Phys.\ B Proc.\ Suppl.\ {\bf#1}, #2 (#3)}
\def \npps#1#2#3{Nucl.\ Phys.\ Proc.\ Suppl.\ {\bf#1}, #2 (#3)}
\def \os{XXX International Conference on High Energy Physics, Osaka, Japan,
July 27 -- August 2, 2000}
\def \PDG{Particle Data Group, D. E. Groom \ite, \epjc{15}{1}{2000}}
\def \pisma#1#2#3#4{Pis'ma Zh.\ Eksp.\ Teor.\ Fiz.\ {\bf#1}, #2 (#3) [JETP
Lett.\ {\bf#1}, #4 (#3)]}
\def \pl#1#2#3{Phys.\ Lett.\ {\bf#1}, #2 (#3)}
\def \pla#1#2#3{Phys.\ Lett.\ A {\bf#1}, #2 (#3)}
\def \plb#1#2#3{Phys.\ Lett.\ B {\bf#1}, #2 (#3)}
\def \pr#1#2#3{Phys.\ Rev.\ {\bf#1}, #2 (#3)}
\def \prc#1#2#3{Phys.\ Rev.\ C {\bf#1}, #2 (#3)}
\def \prd#1#2#3{Phys.\ Rev.\ D {\bf#1}, #2 (#3)}
\def \prl#1#2#3{Phys.\ Rev.\ Lett.\ {\bf#1}, #2 (#3)}
\def \prp#1#2#3{Phys.\ Rep.\ {\bf#1}, #2 (#3)}
\def \ptp#1#2#3{Prog.\ Theor.\ Phys.\ {\bf#1}, #2 (#3)}
\def \rmp#1#2#3{Rev.\ Mod.\ Phys.\ {\bf#1}, #2 (#3)}
\def \rp#1{~~~~~\ldots\ldots{\rm rp~}{#1}~~~~~}
\def \si90{25th International Conference on High Energy Physics, Singapore,
Aug. 2-8, 1990}
\def \slc87{{\it Proceedings of the Salt Lake City Meeting} (Division of
Particles and Fields, American Physical Society, Salt Lake City, Utah, 1987),
ed. by C. DeTar and J. S. Ball (World Scientific, Singapore, 1987)}
\def \slac89{{\it Proceedings of the XIVth International Symposium on
Lepton and Photon Interactions,} Stanford, California, 1989, edited by M.
Riordan (World Scientific, Singapore, 1990)}
\def \smass82{{\it Proceedings of the 1982 DPF Summer Study on Elementary
Particle Physics and Future Facilities}, Snowmass, Colorado, edited by R.
Donaldson, R. Gustafson, and F. Paige (World Scientific, Singapore, 1982)}
\def \smass90{{\it Research Directions for the Decade} (Proceedings of the
1990 Summer Study on High Energy Physics, June 25--July 13, Snowmass, Colorado),
edited by E. L. Berger (World Scientific, Singapore, 1992)}
\def \tasi{{\it Testing the Standard Model} (Proceedings of the 1990
Theoretical Advanced Study Institute in Elementary Particle Physics, Boulder,
Colorado, 3--27 June, 1990), edited by M. Cveti\v{c} and P. Langacker
(World Scientific, Singapore, 1991)}
\def \yaf#1#2#3#4{Yad.\ Fiz.\ {\bf#1}, #2 (#3) [Sov.\ J.\ Nucl.\ Phys.\
{\bf #1}, #4 (#3)]}
\def \zhetf#1#2#3#4#5#6{Zh.\ Eksp.\ Teor.\ Fiz.\ {\bf #1}, #2 (#3) [Sov.\
Phys.\ - JETP {\bf #4}, #5 (#6)]}
\def \zpc#1#2#3{Zeit.\ Phys.\ C {\bf#1}, #2 (#3)}
\def \zpd#1#2#3{Zeit.\ Phys.\ D {\bf#1}, #2 (#3)}

\end{document}